\documentclass[aps,twocolumn,showpacs]{revtex4}
\usepackage{graphicx}

\begin{document}

\title{Effect of interactions on the localization of a Bose-Einstein condensate in a quasi-periodic lattice.}

\author{J. E. Lye$^{1}$, L. Fallani$^{1}$, C. Fort$^{1}$, V. Guarrera$^{1}$, M. Modugno$^{2}$,
D. S. Wiersma$^{1}$, and M. Inguscio$^{1}$}

\affiliation{$^{1}$ LENS, Dipartimento di Fisica and INFM
Universit\`a di Firenze via Nello
Carrara 1, I-50019 Sesto Fiorentino (FI), Italy
\\
$^{2}$ LENS, Dipartimento di Matematica Applicata, Universit\`a di
Firenze and BEC-INFM Center, Universit\`a di Trento, I-38050 Povo
(TN), Italy }


\begin{abstract}
The transport properties of a Bose-Einstein condensate in a 1D
incommensurate bichromatic lattice are investigated both
theoretically and experimentally. We observe a blockage of the
center of mass motion with low atom number, and a return of
 motion when the atom number is increased. Solutions of
the Gross-Pitaevskii equation show how the localization due to the
quasi-disorder introduced by the incommensurate bichromatic
lattice is affected by the interactions.
\end{abstract}



\maketitle

 The intrinsic perfection of lattices made from a standing wave
of light surprisingly makes them an excellent candidate for the
investigation of disorder in atomic systems. Free from
uncontrollable or undesirable defects, precise disorder can be
added simply in the form of additional optical lattices
\cite{Damski,RothandBurnett,Sanchez,Schulte,quasiperiodic,Scarola}
or with an optical speckle potential
\cite{Schulte,Lye,Clement,Fort}. The combination of optical
lattices with a Bose-Einstein condensate (BEC) offers the
stimulating complexity of interactions in a setting unimpeded by
large thermal fluctuations. The significant observation of the
Mott-Insulator phase was realized utilizing a strongly interacting
BEC produced in a three-dimensional ordered crystal of light
\cite{MIone}. Extending this work to the strongly disordered
regime with the inclusion of quasi-disorder from a bichromatic
lattice, recently led to initial experimental evidence of a
Bose-glass phase \cite{boseglass}. Another open question remains
as to the intermediate behavior between the non-interacting
disordered Anderson localized phase and the strongly interacting
Bose-glass phase \cite{logan}.

To begin to address this cross-over regime we have carried out
investigations of the transport properties of a condensate in a
quasi-periodic lattice. The very complexity provided by the
interplay of disorder and interactions that makes disordered
condensates a stimulating topic, also can introduce instability
inherent in non-linear transport \cite{Fallani2004}. In fact we
find that dynamical instability does occur for transport in a
quasi-periodic lattice when the center of mass motion reaches a
critical velocity, and has a non-trivial dependence on the
interaction strength.


Anderson's seminal paper in 1958 showed that the wave function of
a particle placed in a lattice with disordered on-site energies
remains localized when the range of the on-site energies is
sufficiently large compared to the hopping energy between
neighboring sites \cite{Anderson58}.
Anderson's hopping model can be approximated using a
quasi-periodic bichromatic lattice, obtained by superimposing a
primary optical lattice with a weak secondary lattice of a
different incommensurate wavelength. The secondary lattice serves
to break the discrete translational invariance of the system, thus
favoring localization of the wave functions, however the effect of
the quasi \textit{order} may be important depending on the exact
parameters of the bichromatic lattice
\cite{quasiperiodic,Sanchez}.

In our system the 1D incommensurate bichromatic lattice is
produced combining the primary optical lattice derived from a
Titanium:Sapphire laser operating at a wavelength $\lambda_{1} =
830.7(1)$ nm with a secondary lattice obtained from a
fiber-amplified diode laser emitting at $\lambda_{2} = 1076.8(1)$
nm. Our $^{87}$Rb BEC is produced in a Ioffe-Pritchard magnetic
trap, elongated in the direction of the bichromatic lattice. The
trapping frequencies are $\omega_x=2\pi \times 8.7$ Hz axially and
$\omega_\perp=2\pi \times 90$ Hz radially. The BEC can be produced
in the range of $\simeq 1.5 \times 10^4$ - $ 2 \times 10^5$ atoms.
The resulting potential along the lattice axis can be expressed as
\begin{equation}
V(x)=s_{1}E_{R1}\sin^2(k_{1}x)+s_{2}E_{R2}\sin^2(k_{2}x)+\frac{m
}{2}\omega_x^2 x^2
\end{equation}
where  $s_{1}$ and  $s_{2}$  measure the height of the lattice
potentials in units of the respective recoil energies
$E_{R1}=h^2/(2m\lambda_{1}^2)\simeq h \times 3.33$ kHz and $
E_{R2}=h^2/(2m\lambda_{2}^2)\simeq h \times 1.98$ kHz, $k_1$ and
$k_2$ are the wave numbers of the two lasers, $h$ is the Planck
constant and $m$ the mass of $^{87}$Rb.

\begin{figure}[t!]
\begin{center}
\includegraphics[width=0.8\columnwidth]{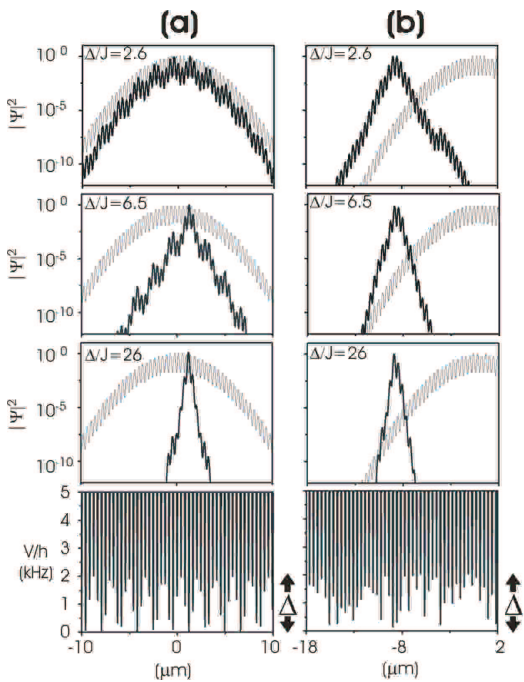}
\end{center}
\caption{Non-interacting density profiles in log scale of the
ground state with increasing disorder in (a) a bichromatic lattice
and (b) a lattice with random on-site energies. The thin line in
each graph represents the ground state with only the primary
lattice. The last graph in each column shows the on-site energies
in the respective potentials with the amount of disorder shown by
$\Delta/J \simeq 26$. In all cases the height of the primary
lattice is $s_1=10$ giving a tunnelling energy of $J/h=75$ Hz.}
\label{setup}
\end{figure}

The possibility of localization with our bichromatic lattice in
the absence of interactions is demonstrated by numerical
diagonalization of the stationary 1D Schr\"{o}dinger equation
using the potential defined by Eq. (1). A strong primary lattice
is chosen, $s_1=10$, $s_1 E_{R1}/h=33$ kHz, and a perturbing
secondary lattice of maximum height $s_2=2$, $s_2 E_{R2}/h=4$ kHz.
The ground state resulting from the bichromatic lattice is shown
in Fig. 1(a) and is contrasted with the ground state of a pure
random case in (b). The random potential is simulated using only
the primary lattice, $s_1=10$, with additional random on-site
energies in a box distribution in the range [0, $\Delta$], where
$\Delta/h \leq 4$ kHz. The amount of the disorder is given by
either the height of the secondary lattice in (a) or by the
maximum on-site energy in the random case (b), and is denoted by
$\Delta$ in both (a) and (b). We define $J$ as the tunnelling in
the primary lattice. The thin line is the ground state with only
the primary lattice, showing the total length of the system in the
harmonic trap. The localized states may have a different center
from the harmonic trap depending on the disorder realization. The
last graph in each column is an enlargement of the actual
potentials. The quasi-periodic system mimics true disorder to a
certain extent, showing localized states characterized by an
exponential decay in the envelope of the density moving away from
the localization center, $|\Psi(x)|^2 \propto \exp(-|x-x_0|/l)$
where $l$ is the localization length. The exponential localization
occurs only above a threshold level at approximately $\Delta/J
\simeq 6$ in the bichromatic case. This is in clear contrast with
random disorder where in a one-dimensional infinite system the
localization behavior persists for any infinitesimal amount of
disorder.


In the presence of weak interactions one expects localization
effects to persist. In fact an Anderson glass phase has been
identified when increasing interactions from an Anderson localized
phase to the Bose-glass state, using the superfluid density as the
order parameter for the phase transition. Unlike the Bose-glass
phase, where both interactions and disorder cooperate to block
diffusion, in the Anderson glass phase the interactions tend to
delocalize the atoms and the disorder must compete against the
interactions \cite{Scalettar91,logan}.

\begin{figure}[t!]
\begin{center}
\includegraphics[width=0.8\columnwidth]{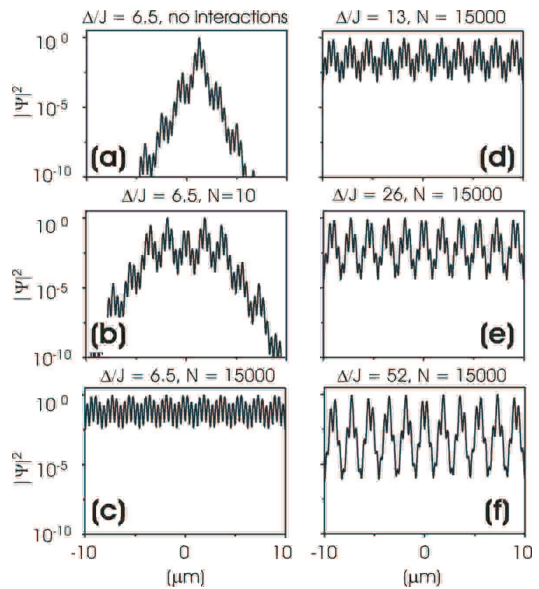}
\end{center}
\caption{Density profile in log scale of the ground state with
interactions for (a)-(c) increasing atom number and fixed
$\Delta/J = 6.5$, and for (d)-(f) increasing $\Delta/J$ and fixed
atom number $N = 1.5 \times 10^4$. In all cases the height of the
primary lattice is $s_1=10$ giving a tunnelling energy of $J/h=75$
Hz.} \label{AG}
\end{figure}

Such behavior can be seen in Fig. 2, showing the ground states in
our incommensurate bichromatic lattice with interactions
calculated by means of a 1D effective Gross-Pitaevskii equation,
namely the nonpolynomial Schr\"{o}dinger equation (NPSE)
\cite{NPSE}.

In the presence of interactions the strongly localized ground
state of (a) transforms into a state with multiple peaks with
partially overlapping tails (b). Upon increasing the interactions,
the overlap between these peaks increases until the state
eventually becomes extended (c). These results show similar
behavior to previous simulations of a BEC in a three-color lattice
\cite{Schulte}. Even with the increased interactions, a state with
multiple peaks can be recovered by increasing disorder (d)-(f).
The cross-over behavior between an extended superfluid state and a
possible Anderson glass state is very difficult to quantify in
this picture. We take a pragmatic approach and characterize the
state of the system by investigating the transport properties.


In the experiment, to set in motion our harmonically-trapped
condensate we excite dipole oscillations by abruptly shifting the
center of the magnetic trap. With a single color lattice, the
superfluid BEC oscillates freely, with the dipole frequency
modified by the effective mass \cite{note1,Cataliotti}. Instead,
adding an incommensurate bichromatic lattice, we expect the
oscillations to be blocked by localization effects. Fig. 3(a)
shows the center of mass motion after a shift in the magnetic trap
of 6 $\mu$m, with a fixed number of atoms $N = 1.5 \times 10^4$, a
fixed height of the primary lattice $s_1 = 10$, and a variable
height of the secondary lattice. At $s_2 = 0.1$ ($\Delta/J=2.6$),
the BEC oscillates with some damping. Increasing $s_2$ the motion
is strongly damped, until at $s_2 = 0.5$ ($\Delta/J=13$), the BEC
does not move. Interestingly, the height of the secondary lattice
at which the condensate becomes localized is slightly greater than
the non-interacting threshold for localization in the bichromatic
lattice shown in Fig 1. The increase of this threshold could be
due to the screening effect of interactions shown in Fig. 2.

We extended the investigations to a variable interaction energy
effectuated by changing the number of atoms in the condensate.
Fig. 3(b) shows the measured center of mass motion of the atoms
for two different atom numbers $N=1.5 \times 10^4$ and $N=2 \times
10^5$, after an initial harmonic trap shift of 6 $\mu$m, $s_1=10$
and $s_2=0.5$ ($\Delta/J=13$). Transport is stopped only for the
low number of atoms, although damping is seen with the higher atom
number.

\begin{figure}[t!]
\begin{center}
\includegraphics[width=0.95\columnwidth]{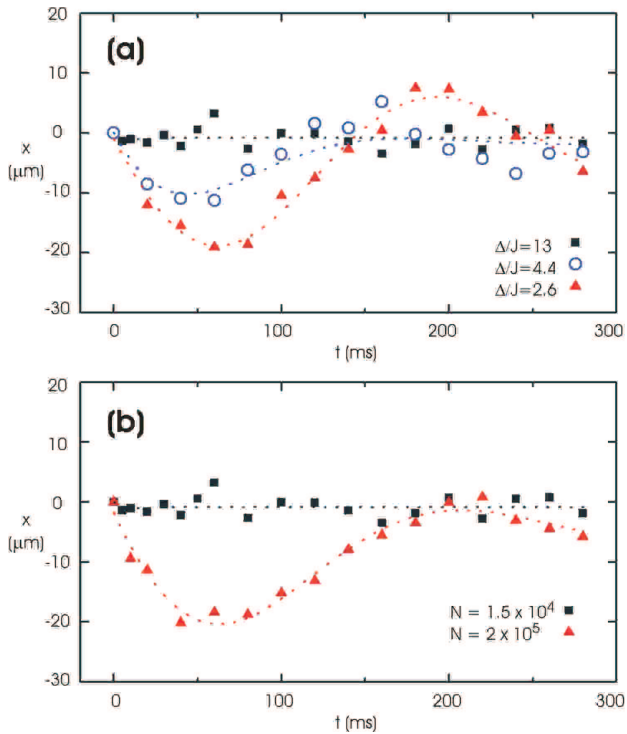}
\end{center}
\caption{(a) Measured dipole oscillations with increasing
intensity of the secondary lattice. The height of the primary
lattice and atom number are fixed at $s_1=10$ and $N=1.5 \times
10^4$ respectively. (b) Measured dipole oscillations with fixed
$s_1=10$, $\Delta/J=13$ and two different atom number, $N=1.5
\times 10^4$ and $N=2 \times 10^5$. The center of mass motion of
the atoms was measured after 20 ms of expansion, fitting to the
central peak. The dotted lines show damped sinusoidals fitted to
the data with the frequency fixed.} \label{blocked}
\end{figure}


Localization due to disorder is not the only physical effect which
can block the motion of the atoms. In the case of an
incommensurate bichromatic lattice the Bloch theorem cannot be
applied, nevertheless the energy spectrum still shows the
existence of energy bands. The band structure for the single
lattice is complicated by the emergence of 'mini-gaps' opening up
almost everywhere across the spectrum \cite{quasiperiodic}.
However the dominant modification to the first Brillouin zone of
the primary lattice, for a weak addition of the secondary lattice,
are the extra energy gaps at $k_{b}$ and $k_1 - k_b$, where
$k_1=2\pi/\lambda_1$ gives the boundary of the first Brillouin
zone of the primary lattice, and $k_b=2\pi/\lambda_b$ corresponds
to the quasi-periodicity introduced on the larger length scale
$\lambda_{b}=\lambda_{1}\lambda_{2}/(\lambda_{2}-\lambda_{1})=4.38\lambda_1$
from the beating between the two colors. This simplification to
the energy spectrum is particularly true when interactions are
introduced, since they can effectively screen the potential
varying on length scales larger than the healing length, washing
away the energy gaps at smaller $k$. As a consequence, in the
presence of interactions one should carefully investigate the
possible contribution of dynamical instability that in a single
lattice has been observed to block dipole oscillations
\cite{Cataliotti, Fallani2004} when the quasi-momentum becomes
greater than  $\simeq 0.5 k_1$. In the bichromatic lattice
dynamical instability may occur at a small quasi-momentum $\gtrsim
0.5 k_{b}$ corresponding to the beat periodicity, much smaller
than the onset of instability with only the primary lattice.

\begin{figure}[t!]
\begin{center}
\includegraphics[width=0.95\columnwidth]{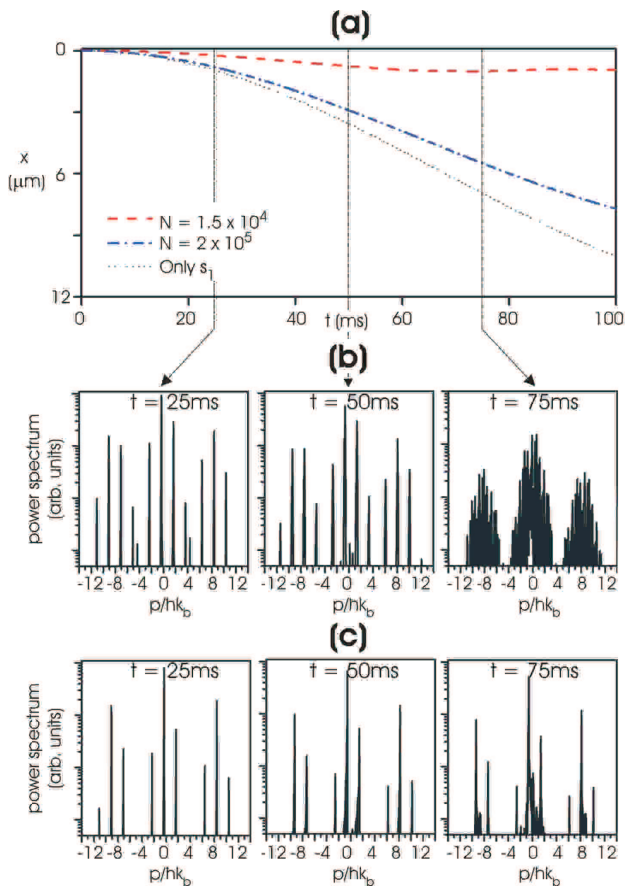}
\end{center}
\caption{(a) The center of mass motion from NPSE simulations of
the dipole oscillations with $s_1=10$, $\Delta/J=13$ and two
different $N$. The dotted line shows the expected superfluid
oscillation in a one color lattice with $s_1 = 10$. (b) The
corresponding momentum spectra in log scale at different times for
$N=1.5\times 10^4$. The vertical arrows show the time at which
each momentum spectrum is taken. (c) The momentum spectra in log
scale at different times for $N=2\times 10^5$.}\label{NPSE}
\end{figure}

To better understand the contribution of the various physical
effects instigating localization, the dipole oscillations in our
actual system were simulated using the time-dependent NPSE. Fig.
4(a) shows the center of mass motion using the same parameters as
the measurements taken in Fig. 3(b); an initial magnetic trap
shift of 6 $\mu$m, $s_1=10$, $s_2=0.5$, and for both $N=1.5 \times
10^4$ and $N=2.0 \times 10^5$. The strongly damped center of mass
motion with low atom number and the increased movement with high
atom number seen in the experimental measurements is confirmed by
the solutions of the NPSE.

From the momentum spectrum it is possible to distinguish when
dynamical instability is present. Fig. 4(b) shows the progression
of the momentum spectra with $N=1.5 \times 10^4$, and Fig. 4(c)
shows the momentum spectra with $N=2 \times 10^5$. At $t=25$ ms,
the main components of the momentum spectra correspond to the
peaks of the primary lattice at integer multiples of $\pm 2 \hbar
k_1$, and peaks of the beat periodicity at integer multiples of
$\pm 2 \hbar k_b$ around the primary peaks. For both high and low
atom number, at $t = 50$ ms extra momentum components begin to
rapidly grow signifying the onset of dynamical instability
\cite{Fallani2004}. We observe that the instability occurs when
the quasimomentum becomes greater than $0.5 k_{b}$. By 75 ms there
is a marked difference between the cases of high and low atom
number. With $N=1.5 \times 10^4$ atoms, the initial spectrum has
been obscured by the additional momentum components. In contrast,
with $N=2 \times 10^5$ atoms, the momentum spectrum retains much
of the original spectrum structure. Intriguingly, the increased
non-linearities inhibit the growth of the instability.

Augmenting the interactions reduces the healing length of the
condensate. In previous experiments, with a single lattice of
spacing 0.4 $\mu$m and typically a healing length of $\simeq 0.3$
$\mu$m, the lattice spacing is not significantly greater than the
healing length, and therefore measurements are largely indifferent
to the atom number \cite{Fallani2004,Cataliotti}.  However, in the
case of the bichromatic lattice the large spacing of the beating
(1.8$\mu$m) allows interactions to effectively smooth over the
large scale beat periodicity. The growth of dynamical instability
in our bichromatic lattice is governed by the competition between
augmentation by increased non-linearity and diminution by
screening of the beat periodicity \cite{note2}.

Consideration of the center of mass motion shown in Fig. 4(a)
together with the momentum spectrum helps to unravel the different
contributions to localization. The contaminated momenta at 75 ms
for low atom number, shown in Fig. 4(b), is reflected in the
complete blockage of the center of mass motion. However, most
importantly, in the absence of dynamical instability, which is
macroscopically observed only after 50 ms, the movement of the
atoms is strongly damped with respect to the superfluid case (see
dotted line in Fig. 4(a)), suggesting that strong damping of the
oscillations is not only due to dynamical instability originating
from the beat periodicity of the bichromatic lattice. The presence
of interactions and the possible screening of the minigaps at
small $k$ from the incommensurate bichromatic lattice renders
interpretation of the blocked motion at short times difficult. As
already pointed out, with sufficiently strong interactions the
energy spectrum could be dominated by the periodicity at
$\lambda_b$, and correspondingly the strongly damped motion at
short times could be described by a high effective mass. Similarly
to the screening behavior seen in Fig. 2, the contribution of the
disorder will depend on the relative strength of interactions and
the intensity of the disordering potential.

In conclusion, maintaining a minimal number of atoms we have
observed a transition from oscillations to blocked motion with
increasing intensity of the incommensurate bichromatic lattice.
Simulations of our experimental parameters using the NPSE show
that the quasi-order inherent in the quasi-periodic bichromatic
lattice leads to the onset of dynamical instability, that
contributes to the blocked motion only after a critical time.
Screening of both localization due to the disorder and dynamical
instability due to the beat periodicity was observed with
strengthened interactions in the simulations. Increasing the
number of atoms in the experiment, we observed a return of
oscillating motion.

This work shows that the exact choice of parameters is crucial to
separate and isolate the effect of disorder-induced localization,
or the non-trivial onset of dynamical instability in a bichromatic
lattice. In future measurements one could focus on measurements at
small quasimomentum to explore the transport behavior without the
complication of dynamical instability, however in an interacting
system the contribution of the disorder to blocked center of mass
motion cannot be easily distinguished from small amplitude
oscillations resulting from an initial high effective mass when
the energy spectrum is dominated by the beat periodicity. In the
non-interacting limit, the bichromatic lattice is also a very
promising tool to investigate Anderson-like localization that
could be accessed utilizing fermions or Feshbach resonances, which
also provide the important possibility of tuning the interactions.

This work has been funded by the EU Contracts No.
HPRN-CT-2000-00125, IST-NoE-Phoremost, MIUR FIRB 2001, MIUR PRIN
2005 and Ente Cassa di Risparmio di Firenze. We thank G. Modugno
for useful discussions and all the LENS Quantum Gases group.



\begin{thebibliography}{99}


\bibitem{Damski}
B. Damski \textit{et al.}, Phys. Rev. Lett. \textbf{91}, 080403
(2003);


\bibitem{RothandBurnett}
R. Roth and K. Burnett, Phys. Rev. A \textbf{68}, 023604 (2003).

\bibitem{Sanchez}
L. Sanchez-Palencia and L. Santos, Phys. Rev. A. \textbf{72}
053607 (2005).

\bibitem{Scarola}
V. W. Scarola and S. Das Sarma, Phys. Rev. A. \textbf{73}
041609(R).

\bibitem{quasiperiodic}
R. B. Diener \textit{et al.}, Phys. Rev. A. \textbf{64} 033416
(2001);

\bibitem{Schulte}
T. Schulte \textit{et al.}, Phys. Rev. Lett. \textbf{95}, 170411
(2005). T. Schulte \textit{et al.}, arXiv:cond-mat/0609774 (2006).


\bibitem{Lye}
J. E. Lye \textit{et al.}, Phys. Rev. Lett. \textbf{95}, 070401
(2005).

\bibitem{Clement}
D. Cl\'{e}ment \textit{et al.}, Phys. Rev. Lett. \textbf{95},
170409 (2005).

\bibitem{Fort}
C. Fort \textit{et al.}, Phys. Rev. Lett. \textbf{95}, 170410
(2005).

\bibitem{MIone}
M. Greiner \textit{et al.}, Nature \textbf{415}, 39 (2002).

\bibitem{boseglass}
L. Fallani \textit{et al.}, arXiv:cond-mat/0603655 (2006)

\bibitem{logan}
P. Lugan \textit{et al.}, arXiv:cond-mat/0610389 (2006)

\bibitem{Fallani2004}
L. Fallani \textit{et al.}, Phys. Rev. Lett. \textbf{93} 140406
(2004).

\bibitem{Anderson58}
P. W. Anderson, Phys. Rev. \textbf{109}, 1492 (1958).

\bibitem{Scalettar91}
R. T. Scalettar, G. G. Batrouni, and G. T. Zimanyi, Phys. Rev.
Lett. \textbf{66}, 3144 (1991).

\bibitem{NPSE}
L. Salasnich, Laser Phys. \textbf{12}, 198 (2002); L. Salasnich,
A. Parola, and L. Reatto, Phys. Rev. A \textbf{65}, 043614 (2002).

\bibitem{note1}
This was confirmed in the experiment for $s_1=10$ and a trap shift
of 6 $\mu$m.

\bibitem{Cataliotti}
F. S. Cataliotti \textit{et al.}, New J. Phys. \textbf{5} 71
(2003)

\bibitem{note2}
A detailed analysis will be presented in a future theoretical
work.





\end{thebibliography}
\end{document}